\documentclass{llncs}
\usepackage{tikz}
\usetikzlibrary{shapes,arrows,calc}

\usepackage{multirow}
\usepackage{listings}
\usepackage{amsmath}  % for equation*
\usepackage{array}    % for tabular
\usepackage{verbatim} % for comment
\usepackage{wrapfig}
\usepackage[final]{microtype}
\usepackage[pdfborder={0 0 0}]{hyperref}
\usepackage{tabularx}

\lstnewenvironment{code}[1][]
  {\noindent
   \vspace{-0.5\baselineskip}
   \lstset{basicstyle=\ttfamily,
           frame=single,
           language=Haskell,
           keywordstyle=\color{black},
           #1}
   \fontsize{8pt}{8pt}\selectfont}
  {}

\newcommand\NOTE[1]{} % \mbox{}\marginpar{\fontsize{8pt}{8pt}\selectfont\raggedright\hspace{0pt}\emph{#1}}}

\begin{document}

\title{Hipster: Integrating Theory Exploration in a Proof Assistant}

\author{Moa Johansson \and Dan Ros\'en \and Nicholas Smallbone \and Koen Claessen}
\institute{Department of Computer Science and Engineering, Chalmers University of Technology
	\email{\{jomoa,danr,nicsma,koen\}@chalmers.se}
	}

\authorrunning{Johansson, Ros\'en, Smallbone, Claessen}
\titlerunning{Integrating Theory Exploration in a Proof Assistant}

\maketitle

\begin{abstract}

This paper describes Hipster, a system integrating theory exploration with the proof assistant Isabelle/HOL. Theory exploration is a technique for automatically discovering new interesting lemmas in a given theory development.
Hipster can be used in two main modes. The first is {\em exploratory mode}, used for automatically generating basic lemmas about a given set of datatypes and functions in a new theory development. The second is {\em proof mode}, used in a particular proof attempt, trying to discover the missing lemmas which would allow the current goal to be proved. Hipster's proof mode complements and boosts existing proof automation techniques that rely on automatically selecting existing lemmas, by inventing new lemmas that need induction to be proved. We show example uses of both modes.
\end{abstract}

% removed from the abstract:
%Existing Isabelle-tools like Sledgehammer can automatically select existing lemmas from the libraries that are likely to be needed for automation of many proofs.

\section{Introduction}
\label{sec:intro}
The concept of theory exploration was first introduced by Buchberger
\cite{buchberger2000theory}. He argues that in contrast to automated
theorem provers that focus on proving one theorem at a time in
isolation, mathematicians instead typically proceed by exploring
entire theories, by conjecturing and proving layers of increasingly
complex propositions. For each layer, appropriate proof methods are
identified, and previously proved lemmas may be used to prove later
conjectures. When a new concept (e.g. a new function) is introduced,
we should prove a set of new conjectures which, ideally, ``completely"
relates the new with the old, after which other propositions in this
layer can be proved easily by ``routine" reasoning. Mathematical
software should be designed to support this workflow. This is arguably
the mode of use supported by many interactive proof assistants, such
as Theorema \cite{theorema} and Isabelle \cite{isabelle}. However,
they leave the generation of new conjectures relating different
concepts largely to the user. Recently, a number of different systems
have been implemented to address the conjecture synthesis aspect of theory exploration
\cite{McCasland2006,isacosy,isascheme,hipspecCADE}. Our work goes one
step further by integrating the discovery and proof of new conjectures
in the workflow of the interactive theorem prover Isabelle/HOL. Our
system, called Hipster, is based on our previous work on HipSpec
\cite{hipspecCADE}, a theory exploration system for Haskell programs.
In that work, we showed that HipSpec is able to
automatically discover many of the kind of equational theorems present
in, for example, Isabelle/HOL's libraries for natural numbers and lists.
In this article we show how similar techniques can be used to speed up and facilitate the
development of new theories in Isabelle/HOL by discovering basic lemmas
automatically.

Hipster translates Isabelle/HOL theories into Haskell and generates equational conjectures by
testing and evaluating the Haskell program. These conjectures are then
imported back into Isabelle and proved automatically. Hipster can be
used in two ways: in \emph{exploratory mode} it quickly discovers
basic properties about a newly defined function and its relationship
to already existing ones. Hipster can also be used in \emph{proof
  mode}, to provide lemma hints for an ongoing proof attempt when the
user is stuck.

Our work complements Sledgehammer \cite{sledgehammer}, a popular Isabelle tool allowing the user to call various external automated provers. Sledgehammer uses \emph{relevance filtering} to select among the available lemmas those likely to be useful for proving a given conjecture \cite{mash}. However, if a crucial lemma is missing, the proof attempt will fail. If theory exploration is employed, we can increase the success rate 
of Isabelle/HOL's automatic tactics with little user effort. 

As an introductory example, we consider the example from section 2.3 of the Isabelle tutorial \cite{isabelle}: proving that reversing a list twice produces the same list. We first apply structural induction on the list \texttt{xs}.
\begin{verbatim}
theorem rev_rev : "rev(rev xs) = xs"
apply (induct xs)
\end{verbatim}
The base case follows trivially from the definition of \texttt{rev},
but Isabelle/HOL's automated tactics \texttt{simp}, \texttt{auto} and
\texttt{sledgehammer} all fail to prove the step case. We can simplify the step case to:
\begin{center}
\texttt{rev(rev  xs) = xs $\Longrightarrow$ rev((rev xs) @ [x]) = x\#xs}
\end{center}
At this point, we are stuck.
This is where Hipster comes into the picture. If we call Hipster at
this point in the proof, asking for lemmas about \texttt{rev} and
append (\texttt{@}), it suggests and proves three lemmas:
\begin{verbatim}
lemma lemma_a:  "xs @ [] = xs"
lemma lemma_aa : "(xs @ ys) @ zs = xs @ (ys @ zs)"
lemma lemma_ab : "(rev xs) @ (rev ys) = rev (ys @ xs)"
\end{verbatim}

To complete the proof of the stuck subgoal, we need lemma \texttt{ab}. Lemma \texttt{ab} in turn, needs lemma \texttt{a} for its base case, and lemma \texttt{aa} for its step case. With these three lemmas present, Isabelle/HOL's tactics can take care of the rest. For example, when we call Sledgehammer in the step case, it suggests a proof by Isabelle/HOL's first-order reasoning tactic \texttt{metis} \cite{metis}, using the relevant function definitions as well as \texttt{lemma\_ab}:
\begin{small}
\begin{verbatim}
theorem rev_rev : "rev(rev xs) = xs"
apply (induct xs)
apply simp
sledgehammer
by (metis rev.simps(1) rev.simps(2) app.simps(1) app.simps(2) lemma_ab)
\end{verbatim}
\end{small}
 
The above example shows how Hipster can be used interactively in a stuck proof attempt. In exploratory mode, there are also advantages of working in an interactive setting. For instance, when dealing with large theories that would otherwise generate a very large search space, the user can instead incrementally explore different relevant sub-theories while avoiding a search space explosion. Lemmas discovered in each sub-theory can be made available when exploring increasingly larger sets of functions. 

The article is organised as follows: In section \ref{sec:background} we give a brief overview of the HipSpec system which Hipster uses to generate conjectures, after which we describe Hipster in more detail in section \ref{sec:hipster}, together with some larger worked examples of how it can be used, both in proof mode and exploratory mode. In section \ref{sec:partial} we describe how we deal with partial functions, as Haskell and Isabelle/HOL differ in their semantics for these. Section \ref{sec:related} covers related work and we discuss future work in section \ref{sec:further}.

\section{Background}
\label{sec:background}

In this section we give a brief overview of the HipSpec system which
we use as a backend for generating conjectures, and of Isabelle's code generator which we use to translate Isabelle theories to Haskell programs. 

\subsection{HipSpec}
HipSpec is a
state-of-the-art inductive theorem prover and theory exploration
system for Haskell. In \cite{hipspecCADE} we showed that HipSpec is
able to automatically discover and prove the kind of equational lemmas present in
Isabelle/HOL's libraries, when given the corresponding functions written in Haskell.

HipSpec works in two stages:
\begin{enumerate}
\item Generate a set of conjectures about the functions at hand. These
  conjectures are equations between terms involving the given
  functions, and have not yet been proved correct but are nevertheless
  extensively tested.

\item Attempt to prove each of the conjectures, using already proven conjectures as assumptions. HipSpec implements this by enumerating induction schemas, and firing off many proof obligations to automated first-order logic theorem provers.
\end{enumerate}
The proving power of HipSpec comes from its capability to
automatically discover and prove lemmas, which are then used to help
subsequent proofs.

In Hipster we can not directly use HipSpec's
proof capabilities (stage (2) above); we use Isabelle/HOL for the proofs instead. Isabelle  is an LCF-style prover which means that it
is based on a small core of trusted axioms, and proofs must be built
on top of those axioms. In other words, we would have to reconstruct
inside Isabelle/HOL any proof that HipSpec found, so it is easier
to use Isabelle/HOL for the proofs in the first place. 

The part of HipSpec we directly use
is its conjecture synthesis system (stage (1) above), called QuickSpec \cite{quickspec}),
which efficiently generates equations about a given set of functions and
datatypes.

QuickSpec takes a set of functions as input, and proceeds to generate all
type-correct terms up to a given limit (usually up to depth three). 
The terms may contain variables (usually at most three per type). 
These parameters are set heuristically, and can be modified by the user. 
QuickSpec attempts to divide the terms into equivalence classes such
that two terms end up in the same equivalence class if they are equal.
It first assumes that all terms of the same
type are equivalent, and initially puts them in the same equivalence class. 
It then picks random ground values for the variables in the terms
(using QuickCheck \cite{quickcheck}) and evaluates the terms.
If two terms in the same equivalence class evaluate to different
ground values, they cannot be equal; QuickSpec thus breaks each equivalence
class into new, smaller equivalence classes depending on what values
their terms evaluated to. This process is repeated until the
equivalence classes stabilise. We then read off equations from each
equivalence class, by picking one term of that class as a
representative and equating all the other terms to that representative.
This means that the conjectures
generated are, although not yet proved, fairly likely to be true, as they have been tested on several hundred different random values. The confidence increases with the number of tests, which can be set by the user. The default setting is to first run 200 tests, after which the process stops if the equivalence classes appear to have stabilised, i.e. if nothing has changed during the last 100 tests. Otherwise, the number of tests are doubled until stable.

As an example, we ask QuickSpec to explore the theory with list append,
\verb~@~, the empty list, \verb~[]~, and three list variables \verb~xs~,
\verb~ys~, \verb~zs~. Among the terms it will generate are
\verb~(xs @ ys) @ zs~, \verb~xs @ (ys @ zs)~, \verb~xs @ []~ and \verb~xs~.
Initially, all four will be assumed to be in the same equivalence class.
The random value generator for lists from QuickCheck might for instance generate the values: \texttt{xs $\mapsto$ []}, \hbox{\texttt{ys $\mapsto$ [a]}} and \texttt{zs $\mapsto$ [b]}, where \texttt{a} and \texttt{b} are arbitrary distinct constants. Performing the substitutions of the variables in the four terms above and evaluating the resulting ground expressions gives us:

\begin{tabularx}{\textwidth}{l  X  X  X}
 & Term & Ground Instance & Value \\
 \hline
1 \quad &\texttt{(xs @ ys) @ zs} & \texttt{([] @ [a]) @ [b]} & \texttt{[a,b]} \\
2 \quad&\texttt{xs @ (ys @ zs)} &\texttt{[] @ ([a] @ [b])} & \texttt{[a,b]}\\
3 \quad&\texttt{xs @ []} & \texttt{[] @ []} & \texttt{[]} \\
4 \quad &\texttt{xs} &\texttt{[]} & \texttt{[]} \\
\end{tabularx}

\noindent Terms 1 and 2 evaluate to the same value, as do terms 3 and 4. The initial equivalence class will therefore be split in two accordingly.
After this, whatever variable assignments QuickSpec generates, the
terms in each class will evaluate to the same value. Eventually, QuickSpec stops and the equations for
associativity and right identity can be extracted from the resulting equivalence classes.

\subsection{Code Generation in Isabelle}
Isabelle/HOL's code generator can translate from Isabelle's higher-order logic to code in several functional programming languages, including Haskell \cite{codegen2,codegen}. Isabelle's higher-order logic is a typed $\lambda$-calculus with polymorphism and type-classes. 
Entities like constants, types and recursive functions are mapped to corresponding entities in the target language. For the kind of theories we consider in this paper, this process is straightforward. However, the code generator also supports user-given \emph{code lemmas}, which allows
it to generate code from non-executable constructs, e.g. by replacing sets with lists.

\section{Hipster: Implementation and Use}
\label{sec:hipster}

\begin{figure}[htbp]
\begin{center}
\includegraphics[scale=0.45]{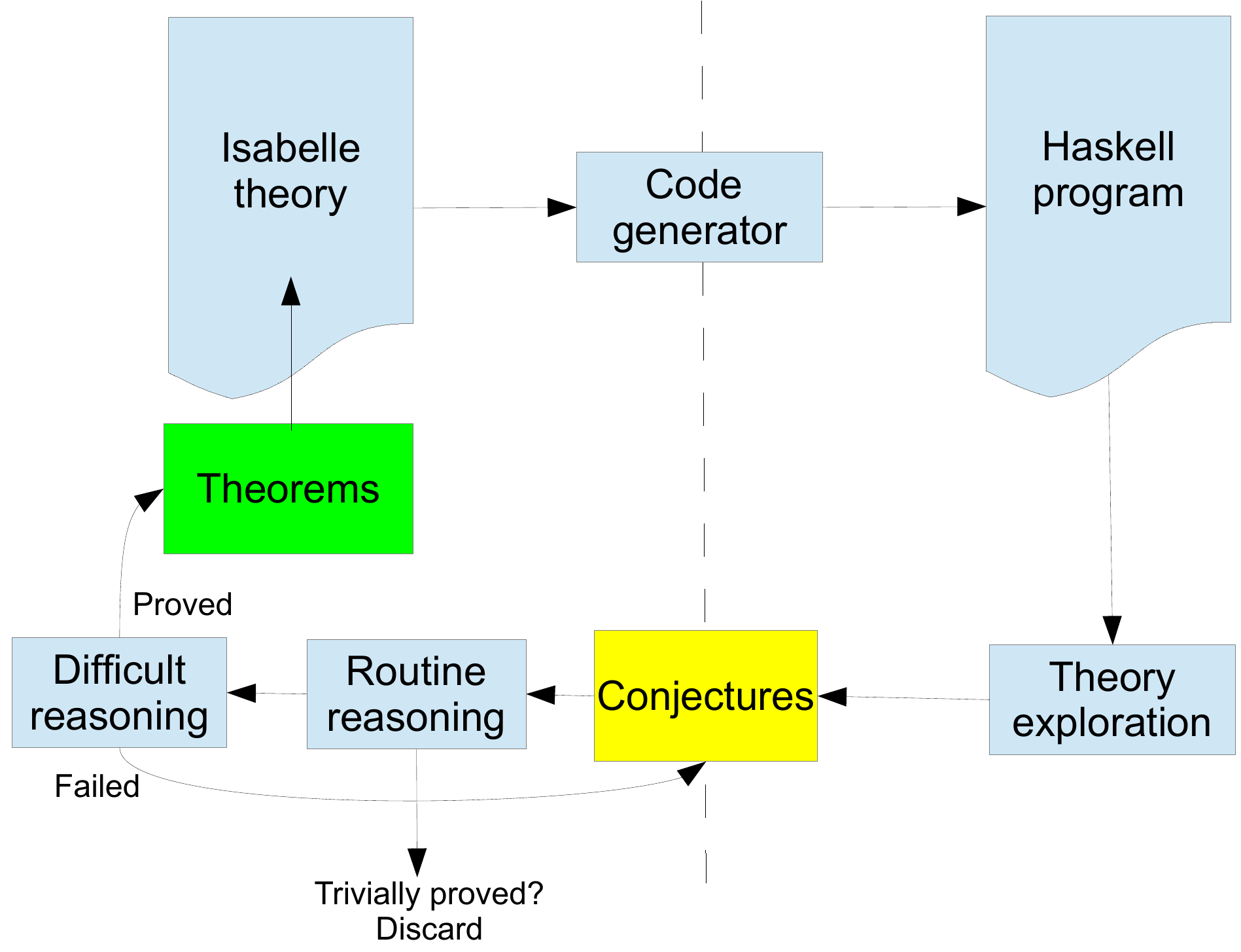}

\caption{Overview of Hipster}
\label{fig:hipster}
\end{center}
\end{figure}

We now give a description of the implementation of Hipster, and show how it can be used both in theory exploration mode and in proof mode, to find lemmas relevant for a particular proof attempt. An overview of Hipster is shown in figure \ref{fig:hipster}. The source code and examples are available online\footnote{\url{https://github.com/moajohansson/IsaHipster}}.

Starting from an Isabelle/HOL theory, Hipster calls Isabelle/HOL's code generator \cite{codegen2,codegen} to translate the given functions into a Haskell program. In order to use the testing framework from QuickCheck, as described in the previous section, we also post-process the Haskell file, adding \emph{generators}, which are responsible for producing arbitrary values for evaluation and testing. A generator in Haskell's QuickCheck is simply a function which returns a random ground value for a particular datatype \cite{quickcheck}. In our case, the generators pick a random constructor and then recursively generate its arguments. To ensure termination, the generators are parametrised by a size; the generator reduces the size when it invokes itself recursively and when the size reaches zero, it always picks a non-recursive constructor. 

Another important issue in the translation is the difference in
semantics for partial functions between Isabelle/HOL and Haskell. In order
for HipSpec not to miss equations that hold in Isabelle/HOL, but not in
Haskell, we have to translate partial functions specially.
This is explained in more detail in section \ref{sec:partial}.

Once the Haskell program is in place, we run theory exploration and generate a set of equational conjectures, which HipSpec orders according to generality. More general equations are preferred, as we expect these to be more widely applicable as lemmas. In previous work on HipSpec, the system would at this stage apply induction on the conjectures and send them off to some external prover. Here, we instead import them back into Isabelle as we wish to produce checkable LCF-style proofs for our conjectures. 

The proof procedure in Hipster is parametrised by two tactics, one for easy or \emph{routine reasoning} and one for \emph{difficult reasoning}. In the examples below, the routine reasoning tactic uses
Isabelle/HOL's simplifier followed by first-order reasoning by Metis
\cite{metis}, while the difficult reasoning tactic performs structural induction followed by simplification and first-order reasoning. 
Metis is restricted to run for at most one second in both the routine and difficult tactic. If there are several possible variables to apply induction on, we may backtrack if the first choice fails. Both tactics have access to the theorems proved so far, and hence get stronger as the proof procedure proceeds through the list of conjectures. 

As theory exploration produces rather many conjectures, we do not want
to present them all to the user. We select the most
interesting ones, i.e. those that are difficult to prove, and filter out those that follow only by routine reasoning.
Depending on the theory and application we can vary the tactics for routine and difficult reasoning to suit our needs. If we want Hipster to produce fewer or more lemmas, we can choose a stronger or weaker tactic, allowing for flexibility.

The order in which Hipster tries to prove things matters. As we
mentioned, it will try more general conjectures first, with the hope
that they will be useful to filter out many more specific routine
results. Occasionally though, a proof will fail as a not-yet-proved
lemma is required. In this case, the failed conjecture is added back
into the list of open conjectures and will be retried later, after at
least one new lemma has been proved. Hipster terminates when it either
runs out of open conjectures, or when it can not make any more
progress, i.e. when all open conjectures have been tried since it last proved a lemma.

Below we give two typical use cases for Hipster. In both examples,
Hipster has been instantiated with the routine and difficult reasoning
tactics that we described above.

\subsection{Exploring a Theory of Binary Trees}
\label{sec:tree}
This example is about a theory of binary trees, with data stored at the leaves:
\begin{small}
\begin{verbatim}
datatype 'a Tree = 
    Leaf 'a 
  | Node "'a Tree" "'a Tree"
\end{verbatim}
\end{small}

Let us define some functions over our trees: \texttt{mirror} swaps the left and right subtrees everywhere, and \texttt{tmap} applies a function to each element in the tree.
\begin{small}
\begin{verbatim}
fun mirror :: 'a Tree => 'a Tree
where
  mirror (Leaf x) = Leaf x
| mirror (Node l r) = Node (mirror r) (mirror l)

fun tmap :: ('a => 'b) => 'a Tree => 'b Tree
where
  tmap f (Leaf x) = Leaf (f x)
| tmap f (Node l r) = Node (tmap f l) (tmap f r) 
\end{verbatim} 
\end{small}
Now, let us call theory exploration to discover some properties about these two functions. Hipster quickly finds and proves the two expected lemmas:  
\begin{small}
\begin{verbatim}
lemma lemma_a [thy_expl]: "mirror (tmap x y) = tmap x (mirror y)"
by (tactic {* Hipster_Tacs.induct_simp_metis . . .*})

lemma lemma_aa [thy_expl]: "mirror (mirror x) = x"
by (tactic {* Hipster_Tacs.induct_simp_metis . . . *})
\end{verbatim}
\end{small}
The output produced by Hipster can be automatically pasted into the proof script by a mouseclick. Recall that Hipster discards all lemmas that can be proved by routine reasoning (here, without induction). The tactic \texttt{induct\_simp\_metis} appearing in the proof script output is the current instantiation of ``difficult reasoning''. Note that the lemmas discovered are tagged with the attribute \texttt{thy\_expl}, which tells Hipster which lemmas it has discovered so far. If theory exploration is called several times, we can use these lemmas in proofs and avoid rediscovering the same things. The user can also inspect what theory exploration has found so far by executing the Isabelle command \texttt{thm thy\_expl}.

Next, let us also define two functions extracting the leftmost and rightmost element of the tree:
\begin{small}
\begin{verbatim}
fun rightmost :: 'a Tree => 'a
where 
  rightmost (Leaf x) = x
| rightmost (Node l r) = rightmost r

fun leftmost :: 'a Tree => 'a
where 
  leftmost (Leaf x) = x
| leftmost (Node l r) = leftmost l
\end{verbatim}
\end{small}
Asking Hipster for lemmas about all the functions defined so far, it provides one additional lemma, namely:
\begin{small}
\begin{verbatim}
lemma lemma_ab [thy_expl]: "leftmost (mirror x2) = rightmost x2"
by (tactic {* Hipster_Tacs.induct_simp_metis . . . *})
\end{verbatim}
\end{small}
Finally, we define a function flattening trees to lists:
\begin{small}
\begin{verbatim}
fun flat_tree :: 'a Tree => 'a list
where
  flat_tree (Leaf x) = [x]
| flat_tree (Node l r) = (flat_tree l) @ (flat_tree r)
\end{verbatim}
\end{small}
We can now ask Hipster to explore the relationships between the functions on trees and the corresponding functions on lists, such as \texttt{rev}, \texttt{map} and \texttt{hd}. Hipster produces four new lemmas and one open conjecture:
\begin{small}
\begin{verbatim}
lemma lemma_ac [thy_expl]: "flat_tree (tmap x y) = map x (flat_tree y)"
by (tactic {* Hipster_Tacs.induct_simp_metis . . . *})

lemma lemma_ad [thy_expl]: "map x (rev xs) = rev (map x xs)"
by (tactic {* Hipster_Tacs.induct_simp_metis . . . *})

lemma lemma_ae [thy_expl]: "flat_tree (mirror x) = rev (flat_tree x)"
by (tactic {* Hipster_Tacs.induct_simp_metis . . . *})

lemma lemma_af [thy_expl]: "hd (xs @ xs) = hd xs"
by (tactic {* Hipster_Tacs.induct_simp_metis . . . *})

lemma unknown: "hd (flat_tree x) = leftmost x"
oops
\end{verbatim}
\end{small}
%flat_tree (tmap x y) = map x (flat_tree y)
%flat_tree (mirror x) = rev (flat_tree x)
%map x (rev xs) = rev (map x xs)
%hd (xs @ xs) = hd xs
Lemmas \texttt{ad} and \texttt{af} are perhaps not of much interest, as they only relate functions on lists. In fact, lemma \texttt{ad} is already in Isabelle/HOL's list library, but is not labelled as a simplification rule, which is why Hipster rediscovers it. Lemma \texttt{af} is a variant of a conditional library-lemma: \hbox{\texttt{xs $\neq$ [] $\Longrightarrow$ hd(xs @ ys) = hd xs}}. Observe that lemma \texttt{af} holds due to the partiality of \texttt{hd}. Hipster can not discover conditional lemmas, so we get this version instead. 

In addition to the four lemmas which have been proved, Hipster also outputs one interesting conjecture (labelled \texttt{unknown}) which it fails to prove. To prove this conjecture, we need a lemma stating that, as the trees store data at the leaves, \texttt{flat\_tree} will always produce a non-empty list: 
 \texttt{flat\_tree t $\neq$ []}. As this is not an equation, it is not discovered by Hipster.
  
This example shows that Hipster can indeed find most of the basic lemmas we would expect in a new theory. The user has to provide the constants Hipster should explore, and the rest is fully automatic, thus speeding up theory development.  Theory exploration in this example takes just a few seconds, no longer than it takes to run tools like Sledgehammer. Even if Hipster fails to prove some properties, they may still be interesting, and the user can choose to prove them interactively.

\subsubsection*{Exploring with different tactics.}
To illustrate the effects of choosing a slightly different tactic for routine and difficult reasoning, we also experimented with an instantiation using only Isabelle/HOL's simplifier as routine reasoning and induction followed by simplification as difficult reasoning. The advantage of this instantiation is that the simplifier generally is faster than Metis, but less powerful. However, for this theory, it turns out that the simplifier is sufficient to prove the same lemmas as above. Hipster also suggests one extra lemma, namely \texttt{rightmost (mirror x) = leftmost x}, which is the dual to lemma \texttt{ab} above. When we used Metis, this lemma could be proved without induction, by routine reasoning, and was thus discarded. Using only the simplifier, difficult reasoning and induction is required to find a proof, and the lemma is therefore presented to the user. 

\subsection{Proving Correctness of a Small Compiler}
\label{sec:comp-ex}
The following example is about a compiler to a stack machine for a toy
expression language\footnote{This example is a slight variation of that in \S3.3 in the Isabelle tutorial \cite{isabelle}. }. We show how theory exploration can be used to unblock a proof on which automated tactics otherwise fail due to a missing lemma.

Expressions in the language are built from constants (\texttt{Cex}), values (\texttt{Vex}) and binary operators (\texttt{Bex}): 
\begin{small}
\begin{verbatim}
type_synonym 'a binop = 'a => 'a => 'a

datatype ('a, 'b) expr =
    Cex 'a 
  | Vex 'b 
  | Bex "'a binop" "('a,'b) expr" "('a,'b) expr"
\end{verbatim}
\end{small}
The types of values and variables are not fixed, but given by type parameters \texttt{'a} and \texttt{'b}.
To evaluate an expression, we define a function \texttt{value}, parametrised by an environment mapping variables to values:
\begin{small}
\begin{verbatim}
fun value :: ('b => 'a) => ('a,'b) expr => 'a
where
    value env (Cex c) = c 
  | value env (Vex v) = env v
  | value env (Bex f e1 e2) = f (value env e1) (value env e2)
\end{verbatim}
\end{small}
A program for our stack machine consists of four instructions:
\begin{small}
\begin{verbatim}
datatype ('a, 'b) program =
    Done
  | Const 'a "('a, 'b) program"
  | Load 'b "('a, 'b) program"
  | Apply "'a binop" "('a, 'b) program"
\end{verbatim}
\end{small}
A program is either empty (\texttt{Done}), or consists of one of the instructions \texttt{Const}, \texttt{Load} or \texttt{Apply}, followed by the remaining program. We further define a function \texttt{sequence} for combining programs:
\begin{small}
\begin{verbatim}
fun sequence :: ('a, 'b) program => ('a, 'b) program => ('a, 'b) program
where
    sequence Done p = p
  | sequence (Const c p) p' = Const c (sequence p p')
  | sequence (Load v p) p' = Load v (sequence p p')
  | sequence (Apply f p) p' = Apply f (sequence p p') 
\end{verbatim}
\end{small}
Program execution is modelled by the function \texttt{exec}, which given a store for variables and a program, returns the values on the stack after execution.
\begin{small}
\begin{verbatim}
fun exec :: ('b => 'a) => ('a,'b) program => 'a list => 'a list
where
    exec env Done stack = stack
  | exec env (Const c p) stack = exec env p (c#stack)
  | exec env (Load v p) stack = exec env p ((env v)#stack)
  | exec env (Apply f p) stack =
     exec env p ((f (hd stack) (hd(tl stack)))#(tl(tl stack)))
\end{verbatim}
\end{small}
We finally define a function \texttt{compile}, which specifies how expressions are compiled into programs:
\begin{small}
\begin{verbatim}
fun compile :: ('a,'b) expr => ('a,'b) program
  where
    compile (Cex c) =  Const c Done
  | compile (Vex v) =  Load v Done
  | compile (Bex f e1 e2) =
     sequence (compile e2) (sequence (compile e1) (Apply f Done))"
\end{verbatim}
\end{small}
Now, we wish to prove correctness of the compiler, namely that executing a compiled expression indeed results in the value of that expression: 
\begin{verbatim}
theorem "exec env (compile e) [] = [value env e]"
\end{verbatim}
If we try to apply induction on \texttt{e}, Isabelle/HOL's simplifier solves the base-case but neither the simplifier or first-order reasoning by Sledgehammer succeeds in proving the step-case. At this stage, we can apply Hipster's theory exploration tactic. It will generate a set of conjectures, and interleave proving these with trying to prove the open sub-goal. Once Hipster succeeds in finding a set of lemmas which allow the open goal to be proved by routine reasoning, it presents the user with a list of lemmas it has proved, in this case:
\begin{small}
\begin{verbatim}
Try first proving lemmas:

lemma lemma_a: "sequence x Done = x"
by (tactic {* Hipster_Tacs.induct_simp_metis . . . *})

lemma lemma_aa: "exec x y (sequence z x1) xs = exec x y x1 (exec x y z xs)"
by (tactic {* Hipster_Tacs.induct_simp_metis . . . *})

lemma lemma_ab: "exec x y (compile z) xs = value x y z # xs"
by (tactic {* Hipster_Tacs.induct_simp_metis . . . *})
\end{verbatim}
\end{small}
Our theorem is a trivial instance of \verb|lemma_ab|, whose proof
depends on \verb|lemma_aa|. Hipster takes about twenty seconds to
discover and prove the lemmas.
Pasting them into our proof script we can try Sledgehammer on our theorem again. This time, it succeeds and suggests the one-line proof:% \texttt{by (metis lemma\_ab)}.
\begin{verbatim}
theorem "exec env (compile e) [] = [value env e]"
by (metis lemma_ab)
\end{verbatim}

% Dealing with Partial Functions
\section{Dealing With Partial Functions}
\label{sec:partial}
Isabelle is a logic of total functions. Nonetheless, we can define
apparently partial functions, such as \verb|hd|:
\begin{verbatim}
fun hd :: 'a list => 'a where
  hd (x#xs) = x
\end{verbatim}

How do we reconcile \verb|hd| being partial with Isabelle functions
being total? The answer is that in Isabelle, \verb|hd| is total, but
the behaviour of \verb|hd []| is unspecified: it returns some
arbitrary value of type \verb|'a|. Meanwhile in Haskell, \verb|head|
is partial, but the behaviour of \verb|head []| is specified: it
crashes. We must therefore translate \emph{partially defined} Isabelle
functions into \emph{total but underspecified} Haskell functions.

Hipster uses a technique suggested by Jasmin Blanchette
\cite{blanchettification} to deal with partial functions. Whenever we translate an Isabelle function
that is missing some cases, we need to add a default case, like so:
\begin{verbatim}
hd :: [a] -> a
hd (x:xs) = x
hd [] = ???
\end{verbatim}

But what should we put for the result of \verb|hd []|? To model the
notion that \verb|hd []| is unspecified, whenever we evaluate a test
case we will pick a \emph{random} value for \verb|hd []|. This value
will vary from test case to test case but will be consistent within
one run of a test case. The idea is that, if an equation involving
\verb|hd| in Haskell always holds, for all values we could pick for \verb|hd []|,
it will also hold in Isabelle, where the value of \verb|hd []| is unspecified.

Suppose we define the function \verb|second|, which returns the second
element of a list, as
\begin{verbatim}
second (x#y#xs) = y
\end{verbatim}
It might seem that we should translate \verb|second|, by analogy with \verb|hd|, as
\begin{verbatim}
second :: [a] -> a
second (x:y:xs) = y
second _ = ???
\end{verbatim}
and pick a random value of type \verb|a| to use in the default case.
But this translation is wrong! If we apply our translated \verb|second|
to a single-element list, it will give the same answer regardless of which
element is in the list, and HipSpec will discover the lemma
\verb|second [x] = second [y]|. This lemma is certainly not true of our
Isabelle function, which says nothing about the behaviour
of \verb|second| on single-element lists, and Hipster will fail to
prove it.

We must allow the default case to produce a different result for
different arguments. We therefore translate \verb|second| as
\begin{verbatim}
second :: [a] -> a
second (x:y:xs) = y
second xs = ??? xs
\end{verbatim}
where \verb|???| is a random \emph{function} of type \verb|[a] -> a|.
(QuickCheck can generate random functions.) As before, whenever we
evaluate a test case, we instantiate \verb|???| with a new random
function\footnote{To avoid having to retranslate the Isabelle theory
every time we evaluate a test case, in reality we parametrise the
generated program on the various \texttt{???} functions. That way,
whenever we evaluate a test case, we can cheaply change the default cases.}.
This second translation mimics Isabelle's semantics: any equation that
holds in Haskell no matter how we instantiate the \verb|???| functions
also holds in Isabelle.

In Hipster, we first use Isabelle/HOL's code generator to translate the
theory to Haskell. Then we transform \emph{every} function definition, whether it is
partial or not, in the same way we transformed \verb|second| above.
If a function is already total, the added case will
simply be unreachable. This avoids having to check functions for partiality.
The extra clutter introduced for total functions is not a problem as we neither reason about nor show the user the generated program.

\section{Related Work}
\label{sec:related}

Hipster is an extension to our previous work on the HipSpec system \cite{hipspecCADE}, which was not designed for use in an interactive setting. HipSpec applies structural induction to conjectures generated by QuickSpec and sends off proof obligations to external first-order provers. Hipster short-circuits the proof part and directly imports the conjectures back into Isabelle. This allows for more flexibility in the choice of tactics employed, by letting the user control what is to be considered routine and difficult reasoning. Being inside Isabelle/HOL provides the possibility to easily record lemmas for future use, perhaps in other theory developments. It gives us the possibility to re-check proofs if required, as well as increased reliability as proofs have been run through Isabelle's trusted kernel. As Hipster uses HipSpec for conjecture generation, any difference in performance (e.g. speed, lemmas proved) will depend only on what prover backend is used by HipSpec and what tactic is used by Hipster. %We remark that some tactics, such as Isabelle's simplifier, can sometimes even make Hipster faster than HipSpec despite running the proofs through Isabelle's , but this depend very much on the theory

There are two other theory exploration systems available for Isabelle/HOL, IsaCoSy \cite{isacosy} and IsaScheme \cite{isascheme}. They differ in the way they generate conjectures, and both discover similar lemmas as HipSpec/Hipster. A comparison between HipSpec, IsaCoSy and IsaScheme can be found in \cite{hipspecCADE}, showing that all three systems manage to find largely the same lemmas on theories about lists and natural numbers. HipSpec does however outperform the other two systems on speed.
IsaCoSy ensures that terms generated are non-trivial to prove by only generating irreducible terms, i.e. conjectures that do not have simple proofs by equational reasoning. These are then filtered through a counter-example checker and passed to IsaPlanner for proof \cite{isaplanner}. IsaScheme, as the name suggests, follows the scheme-based approach first introduced for algorithm synthesis in Theorema \cite{theorema}. IsaScheme uses general user-specified schemas describing the shape of conjectures and then instantiates them with available functions and constants. It combines this with counter-example checking and Knuth-Bendix completion techniques in an attempt to produce a minimal set of lemmas. 

Unfortunately, the counter-example checking in IsaCoSy and IsaScheme
is often too slow for use in an interactive setting. Unlike IsaCoSy,
Hipster may generate reducible terms, but thanks to the equivalence
class reasoning in QuickSpec, testing is much more efficient, and
conjectures with trivial proofs are instead quickly filtered out at
the proof stage. None of our examples takes more than twenty seconds
to run.

Neither IsaCoSy or IsaScheme has been used to generate lemmas in stuck proof attempts, but only in fully automated exploratory mode.
Starting from stuck proof attempts allows us to reduce the size of the
interesting background theory, which promises better scalability.

\emph{Proof planning critics} have been employed to analyse failed
proof attempts in automatic inductive proofs \cite{productiveuse}. The
critics use information from the failure in order to try to speculate
a missing lemma top-down, using techniques based on rippling and
higher-order unification. Hipster (and HipSpec) takes a less
restricted approach, instead constructing lemmas bottom-up, from the symbols available. As was shown in our previous work \cite{hipspecCADE}, this succeeds in finding lemmas that the top-down critics based approach fails to find, at the cost of possibly also finding a few extra lemmas as we saw in the example in section \ref{sec:comp-ex}.

%Hipster has instead been designed specifically to be used interactively during theory development, allowing for more flexibility and user control, both in exploratory and proof mode. For instance, the user can specify what is to be considered difficult or routine reasoning, thus affecting the outcome of the exploration.  

%MATHsAiD, HR?
%Other systems not integrated in the workflow of a proof-assistant. With all the advantages that entail, recording discovered lemmas etc. Building up theories incrementally.

%Unlike HipSpec User can control search space by selecting which functions to pass in together. May run several passes of exploration with different combinations of functions. Parametrised by routine/difficult tactics for flexibility.

\section{Further Work}
\label{sec:further}
The discovery process is currently limited to equational lemmas. We
plan to extend the theory exploration to also take side conditions
into account. If theory exploration is called in the middle of a proof
attempt, there may be assumptions associated with the current
sub-goal, which could be a useful source of side conditions. For
example, if we are proving a lemma about sorting, there will most
likely be an assumption involving the ``less than'' operator; this
suggests that we should look for equations that have ``less than'' in
a side condition. Once we have a candidate set of side conditions, it
is easy to extend QuickSpec to find equations that assume those conditions.

The parameters for Hipster, e.g. the number of QuickSpec generated tests, the runtime for Metis and so on, are largely based on heuristics from development and previous experience. There is probably room for fine-tuning these heuristics and possibly adapt them to the theory we are working in. We plan to add and experiment with additional automated tactics in Hipster. Again, we expect that different tactics will suit different theories.
 
Another interesting area of further work in theory exploration is reasoning by analogy. In the example in section \ref{sec:tree}, theory exploration discovers lemmas about \texttt{mirror} and \texttt{tmap} which are analogous to lemmas about lists and the functions \texttt{rev} and \texttt{map}. Machine learning techniques can be used to identify similar lemmas \cite{acl2ml}, and this information could then be used to for instance suggest new combinations of functions to explore, new connections between theories or directly suggest additional lemmas about trees by analogy to those on lists.

\section{Summary}
\label{sec:concl}

Hipster integrates lemma discovery by theory exploration in the proof
assistant Isabelle/HOL. We demonstrated two typical use cases of how this
can help and speed up theory development: by generating interesting
basic lemmas in \emph{exploration mode} or as a lemma suggestion
mechanism for a stuck proof attempt in \emph{proof mode}. The user can
control what is discovered by varying the background theory, and by
varying Hipster's ``routine reasoning'' and ``difficult reasoning''
tactics; only lemmas that need difficult reasoning (e.g. induction)
are presented.

Hipster complements tools like Sledgehammer: by discovering missing
lemmas, more proofs can be tackled automatically. Hipster succeeds in
automating inductive proofs and lemma discovery for small, but
non-trivial, equational Isabelle/HOL theories. The next step is to
increase its scope, to conditional equations and to bigger theories:
our goal is a practical automated inductive proof tool for Isabelle/HOL.

%Conditionals, analogy between different datastructures when similar lemmas has been found about them. 

\subsubsection*{Acknowledgements} The third author's research was
supported by the EU FP7 Collaborative project {\em PROWESS}, grant
number 317820. The final version of this publication is available at \url{http://link.springer.com}.

\bibliographystyle{plain}
\bibliography{bibfile}

\end{document}